\documentclass[sigconf, nonacm]{acmart}

\usepackage[linesnumbered, ruled, vlined]{algorithm2e}
\SetVlineSkip{0pt}
\SetKwInput{KwInput}{Input}                %
\SetKwInput{KwOutput}{Output}              %
\makeatletter
\patchcmd\algocf@Vline{\vrule}{\vrule \kern-0.4pt}{}{}
\patchcmd\algocf@Vsline{\vrule}{\vrule \kern-0.4pt}{}{}
\makeatother

\usepackage{xcolor}
\usepackage{color, soul}

\usepackage{fancynum}
\usepackage{caption}
\usepackage{subcaption}
\usepackage{graphicx}
\usepackage{array}
\usepackage{tabu}
\usepackage{booktabs}
\usepackage{multirow}
\usepackage{threeparttable}
\usepackage[capitalise,noabbrev]{cleveref}
\usepackage{placeins}

\usepackage[inline]{enumitem}
\setlist{noitemsep,partopsep=0pt,topsep=0pt}
\newlist{inlineenum}{enumerate*}{1}
\setlist[inlineenum]{label=(\roman*),ref=(\roman*)} %
\setlist[itemize,1]{leftmargin=2.2em}

\newsavebox{\figbox}%

\newcommand\vldbdoi{10.14778/3551793.3551829}
\newcommand\vldbpages{2761 - 2773}
\newcommand\vldbvolume{15}
\newcommand\vldbissue{11}
\newcommand\vldbyear{2022}
\newcommand\vldbauthors{\authors}
\newcommand\vldbtitle{\shorttitle} 
\newcommand\vldbavailabilityurl{https://github.com/IDEAL-Lab/shapley-value-independent-utility}
\newcommand\vldbpagestyle{empty} 

\newcommand{\nop}[1]{}

\newtheorem{assumption}{Assumption}
\newtheorem{theorem}{Theorem}
\newtheorem{example}{Example}

\newtheorem{lemma}{Lemma}

\allowdisplaybreaks
\apptocmd\normalsize{%
  \setlength{\abovedisplayskip}{2pt}
  \setlength{\belowdisplayskip}{2pt}
  \setlength{\abovedisplayshortskip}{2pt}
  \setlength{\belowdisplayshortskip}{2pt}
}{}{}

\begin{document}

\title{On Shapley Value in Data Assemblage Under Independent Utility}

 \author{Xuan Luo}
 \affiliation{%
   \institution{Simon Fraser University}
   \streetaddress{8888 University Drive}
   \city{Burnaby}
   \country{BC, Canada}
   \postcode{V5A 1S6}
 }
 \email{xuan_luo@sfu.ca}

\author{Jian Pei}
\affiliation{%
  \institution{Duke University$^1$ and Simon Fraser University$^2$}
  \streetaddress{8888 University Drive}
  \city{$^1$Durham}
  \country{NC, USA and $^2$Burnaby, BC, Canada}
  \postcode{V5A 1S6}
}
\email{j.pei@duke.edu}

 \author{Zicun Cong}
 \affiliation{%
   \institution{Simon Fraser University}
   \streetaddress{8888 University Drive}
   \city{Burnaby}
   \country{BC, Canada}
   \postcode{V5A 1S6}
 }
 \email{zicun_cong@sfu.ca}
 
 \author{Cheng Xu}
 \affiliation{%
   \institution{Simon Fraser University}
   \streetaddress{8888 University Drive}
   \city{Burnaby}
   \country{BC, Canada}
   \postcode{V5A 1S6}
 }
 \email{cheng_xu_3@sfu.ca} 

\nop{
\author{Xuan Luo\textsuperscript{1}, Jian Pei\textsuperscript{1}, Zicun Cong\textsuperscript{1}, and Cheng Xu\textsuperscript{1,2}}
\affiliation{%
  \institution{%
    \textsuperscript{1}Simon Fraser University, Canada\qquad%
    \textsuperscript{2}Hong Kong Baptist University, Hong Kong%
  }
  \institution{xuan\_luo@sfu.ca\qquad jpei@cs.sfu.ca\qquad zicun\_cong@sfu.ca\qquad chengxu@comp.hkbu.edu.hk}
  \country{}
}
}

\begin{abstract}

In many applications, an organization may want to acquire data from many data owners. Data marketplaces allow data owners to produce data assemblage needed by data buyers through coalition.  To encourage coalitions to produce data, it is critical to allocate revenue to data owners in a fair manner according to their contributions.  Although in literature Shapley fairness and alternatives have been well explored to facilitate revenue allocation in data assemblage, computing exact Shapley value for many data owners and large assembled data sets through coalition remains challenging due to the combinatoric nature of Shapley value. In this paper, we explore the decomposability of utility in data assemblage by formulating the independent utility assumption.  We argue that independent utility enjoys many applications. Moreover, we identify interesting properties of independent utility and develop fast computation techniques for exact Shapley value under independent utility. Our experimental results on a series of benchmark data sets show that our new approach not only guarantees the exactness of Shapley value, but also achieves faster computation by orders of magnitudes.

\end{abstract}

\maketitle

\pagestyle{\vldbpagestyle}
\begingroup\small\noindent\raggedright\textbf{PVLDB Reference Format:}\\
\vldbauthors. \vldbtitle. PVLDB, \vldbvolume(\vldbissue): \vldbpages, \vldbyear.\\
\href{https://doi.org/\vldbdoi}{doi:\vldbdoi}
\endgroup
\begingroup
\renewcommand\thefootnote{}\footnote{\noindent
This work is licensed under the Creative Commons BY-NC-ND 4.0 International License. Visit \url{https://creativecommons.org/licenses/by-nc-nd/4.0/} to view a copy of this license. For any use beyond those covered by this license, obtain permission by emailing \href{mailto:info@vldb.org}{info@vldb.org}. Copyright is held by the owner/author(s). Publication rights licensed to the VLDB Endowment. \\
\raggedright Proceedings of the VLDB Endowment, Vol. \vldbvolume, No. \vldbissue\ %
ISSN 2150-8097. \\
\href{https://doi.org/\vldbdoi}{doi:\vldbdoi} \\
}\addtocounter{footnote}{-1}\endgroup

\ifdefempty{\vldbavailabilityurl}{}{
\vspace{.3cm}
\begingroup\small\noindent\raggedright\textbf{PVLDB Artifact Availability:}\\
The source code, data, and/or other artifacts have been made available at \url{\vldbavailabilityurl}.
\endgroup
}

\section{Introduction}
\label{sec:introduction}

The thriving success of data science and machine learning applications heavily relies on the availability of huge amounts of data.  In many applications, an organization may want to empower its business using data but may not have all the necessary data~\cite{DBLP:conf/ec/AgarwalDS19, 10.14778/3447689.3447700}.  At the same time, while organizations can use their own data to strengthen their businesses individually, their data, if being used properly, can help many others, achieve much more social good and bring in dramatic extra value beyond their traditional business.  To facilitate demands and supplies of data meeting each other, various facilities and mechanisms are constructed.  For example, in data marketplaces, organizations and people can buy and sell data~\cite{schomm2013marketplaces, DBLP:journals/pvldb/FernandezSF20, spiekermann2019data}.  The size of data marketplaces over the world has grown dramatically since the last decade, from $7.6$ billion US dollars in 2011 to $64$ billion US dollars in 2021 and $103$ billion US dollars projected in 2027\footnote{\url{https://www.statista.com/statistics/254266/global-big-data-market-forecast/}, accessed on July 1, 2022.}.

The data demand from a buyer may be complicated and thus cannot be met solely by one data owner.  
Such data may have to come from many different data owners.  
To meet the complicated demand, data from multiple owners has to be integrated and assembled. 
To encourage data owners to produce valuable data for many applications through coalition, it is critical to allocate revenue to data owners in a fair manner according to their contributions. However, fair revenue allocation in data assemblage is far from trivial.  The celebrated Shapley fairness~\cite{Shapley} is the most fundamental and popular fairness principle used in marketplaces.
In a coalition, the Shapley value of a participant is essentially the expectation of marginal contribution made by the participant in all possible coalitions with various subsets of other participants. Shapley value enjoys a series of desirable properties, including efficiency in revenue allocation, symmetry, additivity, and dummy player. %

Due to the combinatoric nature of Shapley value, computing the exact Shapley value is often very costly and in general is exponential with respect to the number of participants in coalition~\cite{ghorbani2019data, jia2019towards}.  Therefore, approximation approaches are developed.  In those approximation approaches, the estimation error can be bounded by, for example $O(\sqrt{\frac r m})$~\cite{maleki2013bounding}, where $r$ is the range of marginal contributions and $m$ is the sample size.  While mathematically the estimation error bound may look satisfying, in practice the approximation quality of Shapley value may be much less impressive.  Consider the scenarios of data acquisition through crowdsourcing, where there may easily be tens of thousands of data owners contributing their data.  Due to the diversity of data contributors, the contributions from different data owners may also vary dramatically and often demonstrate a long tail distribution.  Therefore, the range of marginal contributions is significant, say easily much greater than $1\%$ of the total revenue of the coalition.  Even if the estimation error can be bounded to a small percentage, say $0.1\%$, of the whole revenue to be allocated, the absolute error of the estimated Shapley values of those participants not in the head may easily larger than their true Shapley values, as demonstrated by our experiments, too.  In such situations, exact Shapley values are highly desirable.

Moreover, even estimating Shapely value may still be time consuming.  When there are many data owners, a large number of samples are needed to accomplish a reasonable estimation.  In general, in order to achieve an error bound $\epsilon$ (in percentage of the total revenue to be allocated) with probability at least $1-\alpha$, that is, to ensure $P(\|\widehat{\psi(u)}-\psi(u)\| \leq \epsilon)\geq (1-\alpha)$, we need at least $O(\frac{Z_{\alpha/2}^2\sigma^2}{\epsilon^2})$ samples, where $\psi(u)$ and $\widehat{\psi(u)}$ are the Shapley value and the estimation, respectively, $Z_{\alpha/2}$ is the value such that $P(Z \geq Z_{\alpha/2})=\alpha/2$, $\sigma^2$ is the sample variance, and $Z \sim N(0, 1)$~\cite{CASTRO20091726}.  As shown in our experiments, the Monte Carlo simulation costs dramatic time when there are many data owners in assemblage of large data sets.

Most of the existing studies in Shapley value computation for data marketplaces assume general utility functions, such as accuracy of machine learning models and database query results~\cite{ghorbani2019data, jia2019towards, jia2019efficient}.  We observe that, due to the fine granularity and powerful composability of data, in many applications, utility in data assemblage has some unique and useful properties.  First, as pointed out by established research in data economics~\cite{1998information, 9300226}, the basic units in data are often well defined and fixed.  For example, when an organization wants to acquire data for marketing, each customer record is a unit and the utility of the record often can be evaluated independently in business.   This is very different from the conventional physical products where the utility of a part by itself is almost useless. Second, data records often have strong composability. For example, in many situations, the utility of a collection of customer records for marketing can be estimated well by the sum of the utility of the individual records~\cite{DBLP:journals/pvldb/UpadhyayaBS16}. The decomposability not only is intuitive, but also facilitates business operations foundamentally.

Can we explore the decomposability of utility in data assemblage and achieve exact and fast Shapley value computation?  This insight motivates our study here.  Based on the above analysis, we develop a simple and intuitive independent utility assumption -- the utility of a unit in the result of coalition can be assessed independently and the total utility of the coalition is the sum of the utility of all units produced in the coalition. This independent utility assumption holds in many applications where data is assembled by coalition among multiple data owners and provided to data buyers for consumption, such as many demands and supplies of survey data, micro data, and review data in data marketplaces. For example, data marketplaces like Windows Azure Marketplace\footnote{\url{https://azuremarketplace.microsoft.com/}, accessed on May 3, 2022.} support pricing an API call by summing up the cost of each tuple returned by the call. In data marketplace Datarade\footnote{\url{https://datarade.ai/},  accessed on May 3, 2022.}, there are multiple data vendors like Traject Data\footnote{\url{https://datarade.ai/data-products/local-business-reviews-traject-data}, accessed on May 3, 2022.} and Mapping Resources \footnote{\url{https://datarade.ai/data-products/united-states-individual-and-household-consumer-list-database-mapping-resources}, accessed on May 3, 2022.} that charge per record.

To the best of our knowledge, we are the first to explore independent utility and utility decomposability in general for data coalition.
Under independent utility how can we compute Shapley value fast?
There are a series of challenges. 
Straightforwardly applying the independent utility assumption to the existing methods may not reduce the computation cost. Existing methods compute the exact Shapley value of a participant by considering the marginal contributions made by the participant in all possible coalitions with other participants. For every possible coalition, it is needed to assemble data from all participants in a coalition and then calculate the total utility.  The existing approximation methods may rely on Monte Carlo simulation and demand a large number of samples to reduce the error to a sufficient level. 
The independent utility assumption cannot help the existing methods to reduce the number of coalitions, the workload of assembling in each coalition, or the required sample size.
Moreover, when we explore the decomposability of Shapley value to reduce the cost in an exponential number of coalitions and in data assemblage, one data record may still be produced by multiple data owners in different ways. The combinatoric nature of Shapley value remains at the unit level. 

To tackle the challenges, we systematically analyze the possible situations where a record in the coalition results is produced and develop corresponding methods, some with closed form and some with fast algorithms.
Importantly, we investigate the interesting tradeoffs in computation cost between the number of contributing data owners and the number of possible ways a record is produced.  Our overall approach smartly chooses the faster way to derive exact Shapley value according to how a result record is produced.

To evaluate the effectiveness and efficiency of our approach, we conduct extensive experiments on a series of benchmark data sets and compare with the state-of-the-art exact Shapley value computation baseline and Monte Carlo based baselines.  The results clearly show that our new approach not only can always guarantee the exactness of Shapley value but also can achieve faster computation by orders of magnitudes.

The rest of the paper is organized as follows.  We review the related work in Section~\ref{sec:related-work}.  Then, we formulate the independent utility assumption in Section~\ref{sec:prof_defn}. We explore the decomposability of Shapley value computation under independent utility and tackle two basic cases in Section~\ref{sec:horizontal}.
We tackle the general situation and complete our method in Section~\ref{sec:algo}.  We report the experimental results in Section~\ref{sec:exp} and conclude the paper in Section~\ref{sec:con}.  

\section{Related Work}
\label{sec:related-work}

Due to the strong rise of data science, many %
data marketplaces~\cite{schomm2013marketplaces, DBLP:journals/pvldb/FernandezSF20} are constructed, where data demands and supplies can meet.  Some examples of data marketplaces include Dawex\footnote{\url{https://www.dawex.com/en/}, accessed on May 9, 2021.}, Snowflake data marketplace\footnote{\url{https://www.snowflake.com/data-marketplace/}, accessed on May 9, 2021.}, and BDEX\footnote{\url{https://www.bdex.com}, accessed on May 9, 2021.}. 
\citet{muschalle2012pricing} identify seven categories of participants in data marketplaces.
In addition to data marketplaces, model marketplaces, where the acquisition of data is mainly to train machine learning models, is also an emerging research topic~\cite{10.14778/3447689.3447700, DBLP:conf/sigmod/ChenK019}. While many common principles are applicable to both data marketplaces and model marketplaces~\cite{9300226}, a critical difference is on the utility measure.  In model marketplaces, utility is often measured using model performance, such as accuracy and precision.  General data marketplaces do not hold such an assumption about the utility measure. For data pricing for machine learning tasks, \citet{DBLP:journals/kais/CongLPZZ22} provide a survey. In this paper, we focus on data marketplaces.

In data and model marketplaces, a critical issue is fair revenue allocation.  \citet{Shapley} establishes the Shapley fairness, which is the most fundamental and popular fairness principle used in marketplaces (see Section~\ref{sec:Shapley} for a brief review). Computing Shapley value is often costly due to its combinatoric nature.  Some alternatives are, for example, leave-one-out~\cite{doi:10.1080/00401706.1980.10486199}, which measures the value of a data point by the difference between the utility of the whole data set and the set leaving the data point out.  It does not satisfy all the four desirable properties of Shapley fairness and may not properly evaluate the value of a data point.  For example, in a data set where there are two identical data points $u$ and $v$, leave-one-out assigns them value $0$ since leaving one out does not affect the utility of the rest of the data set~\cite{DBLP:conf/icml/YoonAP20}.

To tackle the challenges in computing Shapley value,~\citet{maleki2013bounding} propose the Monte Carlo approximation method.  Assuming more properties of utility functions in marketplaces facilitates further approximations.  For example,~\citet{kleinberg2001value} consider the utility of a coalition as the number of unique items in the assembled data sets of the coalition and thus the Shapley value of a data owner is the total ``novelty'' of the owner's data items. The novelty of an item is inversely proportional to the number of data owners having the item. \citet{ghorbani2019data} assume that, in supervised learning, the performance change of a model may be ignorable if only one or very few training data points are added. They  develop the truncated-based and gradient-based approximation methods for Shapley values of individual data points.  \citet{jia2019towards} approximate Shapley value using group testing.  \citet{jia2019efficient} exploit locality of utility in some models, such as $k$-nearest neighbor classifiers, and develop polynomial time complexity methods.

Our study is different from the previous methods in two aspects. First, in this paper, we consider the situations where the utility of a record in the coalition result set is independent from the other records.  Independent utility is a fundamental and natural property that rises in many applications. To the best of our knowledge, the independent utility and decomposability of utility in coalition are not considered by any previous studies.
Second, we systematically explore the opportunity of efficient computation of exact Shapley value enabled by independent utility.  Most of the previous methods on fast computation of Shapley value have to adopt approximation.

Our study is also remotely related to pricing database queries~\cite{DBLP:conf/pods/KoutrisUBHS12, DBLP:conf/sigmod/KoutrisUBHS13, DBLP:conf/icdt/DeepK17, DBLP:conf/sigmod/DeepK17, DBLP:journals/pvldb/ChawlaDKT19, miao2020towards,DBLP:journals/pvldb/UpadhyayaBS16}. A critical difference is that pricing database queries tries to set a price for a query extracting information from a database, while our study computes the Shapley value of every data owner who contributes to a coalition. In pricing database queries, there is only one data owner but there are many queries that can be regarded as data buyers. In our study, we consider only one data buyer but many data owners.  Therefore, those methods for pricing database queries cannot be used to tackle the problem studied in this paper.  Although \citet{DBLP:conf/sigmod/KoutrisUBHS13} consider revenue sharing among possibly multiple data owners in pricing database queries, their revenue allocation method does not satisfy the Shapley fairness requirement.

Finally, our study is also related to quantifying the contribution of database tuples to query answers~\cite{DBLP:conf/icdt/LivshitsBKS20, DBLP:journals/pvldb/MeliouGMS11, DBLP:journals/pvldb/MeliouRS14, DBLP:conf/tapp/SalimiBSB16, DBLP:journals/corr/abs-2112-08874}. The major difference is that those methods mainly focus on numerical queries that map databases to numbers, like Boolean queries, %
but our study puts no constraint on query types and can be applied to any queries in general. %

\section{Problem Formulation}
\label{sec:prof_defn}

In this paper, we consider \textbf{data assemblage} from a set of \textbf{data owners} $\mathcal{U}=\{u_1, \ldots, u_n\}$.  Each data owner $u_i$ $(1 \leq i \leq n)$ owns a data set.  For the sake of clarity, we overload the symbol $u_i$ to denote the data set owned by owner $u_i$. A \textbf{buyer} wants to get as much data as possible from the data owners that meets the buyer's need.  The data owners also want to contribute their data to produce as many unique data records\footnote{In this paper, we use the terms ``record'' and ``tuple'' interchangeably.} as possible meeting the buyer's need.  A \textbf{coalition plan}, denoted by $\mathcal{P}$, that is, how the data sets from data owners are used to assemble tuples in the outcome of coalition, is specified and managed by a broker or coordinator.  The data set $D$ produced by coalition is called the \textbf{coalition (data) set}. We assume that there are no duplicate tuples in $D$. That is, although multiple combinations of data owners may produce the same tuple in the coalition set, those duplicates are combined into one.  The \textbf{data assemblage task} is to assemble the data for the buyer by coalition among the data owners.

\begin{example}[Data assemblage]\label{ex:assemblage}\rm
Suppose a data buyer wants to collect data about potential customers and companies in schema $R=($customer-pseudo-identifier, company$)$.  A social media $u_1$ as a data owner may provide data in schema $R_1=($customer-pseudo-identifier, product$)$, which records the potential interest on products by customers.  Another social media $u_2$ as a data owner may provide data in schema $R_2=($customer-pseudo-identifier, brand$)$, which records the customers' interest on brands.  A data integration company $u_3$ provides the mapping among products, brands, and companies in schema $R_3=($product, brand, company$)$.  

Suppose the tables of the data owners are, respectively, $u_1=\{(\#10093, \mbox{911 Targa})\}$,
$u_2=\{(\#10093$, $\mbox{Audi})\}$, and
$u_3=\{(\mbox{A6}$, $\mbox{Audi}$, $\mbox{Volkswagen})$, $\ldots$, $(\mbox{911 Targa}$, $\mbox{Porsche}$, $\mbox{Volkswagen})\}$.  Here, we overload the symbols $u_1$, $u_2$, and $u_3$ to denote the data sets of the three owners, respectively.

To produce the data in the coalition set, a broker/coordinator specifies the coalition plan 
\begin{align*}
\mathcal{P}= & \mbox{\emph{Proj}}_{\mbox{customer-pseudo-identifier, company}}(u_1 \bowtie u_3) \\
&\cup \mbox{\emph{Proj}}_{\mbox{customer-pseudo-identifier, company}}(u_2 \bowtie u_3)
\end{align*}
where \emph{Proj} and $\bowtie$ are the projection and natural join operations on relational data, respectively\footnote{We reserve symbol $\Pi$ for the set of all permutations.}.

The coalition set $D$ contains only one tuple $(\#10093$, $\mbox{Volkswagen})$, which is generated by the coalition between $u_1$ and $u_3$, as well as that between $u_2$ and $u_3$.  Indeed, $u_1$ and $u_3$ together produce an instance and $u_2$ and $u_3$ together produce another instance, but the coalition set merges the two duplicates into one.
\qed
\end{example}

\subsection{Shapley Fairness and Shapley Value}\label{sec:Shapley}

If a buyer pays a reward $v$ for the data produced by the coalition among the data owners, how can we distribute the reward as the revenue\footnote{In this paper, the terms ``reward'' and ``revenue'' are used interchangeably.} to the owners in a fair way to properly recognize their contributions? Shapley~\cite{Shapley} establishes the \textbf{Shapley fairness} in recognizing the individual contribution in a coalition.  Let $v_i$ $(1 \leq i \leq n)$ be the payment to data owner $u_i$. There are four fundamental requirements to achieve a fair allocation.

\begin{itemize}

\item \textbf{Balance.} The payment of $v$ is fully distributed to all data owners, that is $\sum_{i=1}^n v_i = v$.  This property is also known as efficiency of revenue allocation.

\item \textbf{Symmetry.}  The same contributions lead to the same payment.  Given two data owners $u_i$ and $u_j$ $(1 \leq i, j \leq n)$, if for every subset of data owners $\mathcal{S} \subset \mathcal{U}$ such that $u_i \not\in \mathcal{S}$ and $u_j \not\in \mathcal{S}$, the utility of $\mathcal{S} \cup \{u_i\}$ and that of $\mathcal{S} \cup \{u_j\}$ are the same, then $v_i =v_j$.

\item \textbf{Zero element.} No contribution, no payment. For data owner $u_i$, if for every subset of data owners $\mathcal{S} \subset \mathcal{U}$ such that $u_i \not\in \mathcal{S}$, the utility of $\mathcal{S} \cup \{u_i\}$ and that of $\mathcal{S}$ are identical, then $v_i=0$.  This property is also known as null player.

\item \textbf{Additivity.}  If the coalition can be used for two tasks and thus two payments $v$ and $v'$ are obtained, then the payment to complete both tasks is $v + v'$.  This property is also called the additivity.

\end{itemize}

In the above well celebrated Shapley fairness, the \textbf{Shapley value} is the unique allocation of payment that satisfies all the requirements. For any data owner $u \in \mathcal{U}$, the Shapley value of $u$ is 
\begin{equation}\label{eq:shapley}
\psi(u) = \frac 1 {\|\mathcal{U}\|} \sum_{\mathcal{S} \subseteq \mathcal{U} \setminus \{u\}} \frac{Utility(\mathcal{S} \cup \{u\})- Utility(\mathcal{S})}{\binom{n-1} {\|\mathcal{S}\|}}
\end{equation}
where $Utility(\cdot)$ is a utility function and $\mathcal{U}$ is the complete set of data owners.  For the sake of simplicity, we overload the function $Utility(\cdot)$ so that it can take either a set of data owners or a set of data records owned by multiple data owners as input, and returns the utility of the coalition among all the data owners and the utility of the coalition using all the data, respectively.

Equivalently, Equation~\ref{eq:shapley} can also be rewritten as
\begin{equation}\label{eq:shapley-permutation}
\psi(u)=\frac 1 {\|\mathcal{U}\|!} \sum_{\pi \in \Pi(\mathcal{U})}(Utility(P^\pi_u\cup\{u\})-Utility(P^\pi_u))
\end{equation}
where $\Pi(\mathcal{U})$ is the set of all possible permutations of all data owners, $\pi$ is a permutation, and $P_u^\pi$ is the set of data owners preceding $u$ in $\pi$.

Computing Shapley value using Equations~\ref{eq:shapley} and~\ref{eq:shapley-permutation} is often very costly and cannot scale up to a large set of data owners, due to the combinatorial nature of the problem.  

\subsection{Independent Utility}

In the context of acquiring records in a database, the basic units are often records.  As illustrated in Section~\ref{sec:introduction}, in many applications, it is natural and reasonable to assume that the utility of a set of tuples is the sum of the utility of individual tuples in the set, and the utility of tuples is independent from each other.  We formalize the notion of independent utility.

\begin{assumption}[Independent Utility]
The \textbf{independent utility assumption} holds on a data set $D =\{t_1, \ldots, t_l\}$ if the utility of the data set $Utility(D)=\sum_{i=1}^l Utility(t_i)$ and for any $1 \leq i, j \leq l$, $Utility(t_i)$ and $Utility(t_j)$ are non-negative and independent from each other.
\end{assumption}

In this paper, for the ease of presentation, most of the time we assume each tuple has the same utility $1$.  However, our discussion can be straightforwardly extended to entertain the scenarios where different tuples may carry different utility values.

Under the independent utility assumption, the Shapley value of a data owner with respect to a coalition set can be decomposed into the Shapley value of the data owner with respect to every individual tuple in the coalition set. 

\begin{theorem}[Independent Shapley Value]\label{thm:independent}
Let $D=\{t_1, \ldots, t_l\}$ be a coalition set produced by a coalition by data owners $\mathcal{U}=\{u_1, \ldots, u_n\}$. Under the independent utility assumption, for every data owner $u_i$ $(1 \leq i \leq n)$, the Shapley value of $u_i$ is $\psi(u_i)=\sum_{j=1}^l \psi_{t_j}(u_i)$, where $\psi_{t_j}(u_i)$ is the Shapley value of $u_i$ in producing tuple $t_j$ by coalition.%

\proof 
Due to the independent utility assumption, for any subset of tuples $D' \subseteq D$, $Utility(D')=\sum_{t \in D'} Utility(t)$. According to Equation~\ref{eq:shapley-permutation},
\begin{align*}
\psi(u_i)  =& \frac 1 {n!} \sum_{\pi \in \Pi(\mathcal{U})}(Utility(P_{u_i}^\pi \cup \{u_i\})-Utility(P_{u_i}^\pi))\\
=& \frac 1 {n!} \sum_{\pi \in \Pi(\mathcal{U})}\sum_{j=1}^l ( Utility(t_j) \cdot I(t_j \in P_{u_i}^\pi \cup \{u_i\}) \\
&   - Utility(t_j) \cdot I(t_j \in P_{u_i}^\pi) )\\
=&  \sum_{j=1}^l (\frac 1 {n!} \sum_{\pi \in \Pi(\mathcal{U})} [ Utility(t_j) \cdot I(t_j \in P_{u_i}^\pi \cup \{u_i\}) \\
&   - Utility(t_j) \cdot I(t_j \in P_{u_i}^\pi) ] )\\
=& \sum_{j=1}^l \psi_{t_j}(u_i)
\end{align*}
where $I(\cdot)$ is the indicator function, and $t_j \in \mathcal{S}$ is overloaded to denote that tuple $t_j$ can be produced by a coalition among the data owners in subset $\mathcal{S} \subseteq \mathcal{U}$.
\qed
\end{theorem}

Theorem~\ref{thm:independent} indicates that, for each data owner, the Shapley value with respect to individual data tuples are also independent from those of other data tuples that may be produced by coalition.  This nice property of decomposability enables new opportunities to calculate Shapley value efficiently in data assemblage.

For a set of data owners $\mathcal{S}$ who try to produce a tuple $t \in D$ in the coalition set through coalition, we write $Utility_t(\mathcal{S})$ to represent the utility with respect to $t$.  When $\mathcal{S}$ can produce $t$, $Utility_t(\mathcal{S})=Utility(t)$; otherwise, $Utility_t(\mathcal{S})=0$.

\subsection{Complexity and Opportunities}

Unfortunately and not surprisingly, even under independent utility, Shapley value computation is still NP-hard. This can be easily shown by a reduction from the Shapley value computation in (weighted) $k$-majority games~\cite{Osborne1994, RePEc:eee:ejores:v:143:y:2002:i:1:p:181-196, DBLP:journals/ai/FatimaWJ08}.

\nop{
\begin{theorem}[Complexity]
Shapley value computation under the independent utility assumption is NP-hard.

\proof
We prove by a reduction from the Shapley value computation in (weighted) $k$-majority games~\cite{Osborne1994, RePEc:eee:ejores:v:143:y:2002:i:1:p:181-196, DBLP:journals/ai/FatimaWJ08}. Clearly, since a weighted $k$-majority game consists of $k$ independent weighted voting games, it naturally satisfies the independent utility assumption.  Moreover, for the special case $k=1$, it is a weighted voting game, where the computation of Shapley value is \#P-hard~\cite{10.2307/3690220}. 
\qed
\end{theorem}
}

The intractability of Shapley value under independent utility does not prevent us from exploring fast methods in practice.  Particularly, under independent utility, sparsity provides significant opportunities.  In the context of data assemblage, although there may be many data owners and many tuples in a coalition set, for one specific tuple in the coalition set, there are typically very few ways to produce the tuple and very few data owners involved.  Exactly due to this sparsity and scarcity, data pricing becomes meaningful.  To this extent, our proposal on independent utility is a step towards exploring efficient data pricing addressing the scarcity of supplies of specific data.

\section{Syntheses and Shapley Value Computation Decomposition}\label{sec:horizontal}

In order to compute Shapley value, we need to model how tuples in a coalition set are synthesized by data owners according to the coalition plan.  Under independent utility, can the computation of Shapley value of one tuple in the coalition set also be decomposed according to syntheses? In this section, we first propose the notion of synthesis.  Then, we answer the above question in two simple cases.  In the first case, a tuple in the coalition set is contributed by data owners independently without any interaction, that is, there is only one owner in a synthesis.  In the second case, a tuple in the coalition set is produced by multiple data owners together, but there is only one way to produce an instance of the tuple by multiple data owners.  For both cases, we give closed form solutions.

\subsection{Syntheses}

For a tuple in the coalition set $t \in D$, if data owners $u_{i_1}, \ldots, u_{i_m}$ in coalition produce an instance of $t$ according to the coalition plan $\mathcal{P}$, then $U=\{u_{i_1}, \ldots, u_{i_m}\}$ is called a \textbf{synthesis} of $t$. As a special case, a data owner can produce a tuple in the coalition set by itself if the owner has the complete tuple.  Trivially, the corresponding synthesis has only one data owner.  We call a synthesis $U=\{u_{i_j}\}$ that contains only one data owner a \textbf{single-owner synthesis}.  
\nop{
\begin{example}[Single-owner syntheses]\label{ex:single-owner}\rm
Consider 3 data owners, $u_1$, $u_2$, and $u_3$.  Suppose the data owners' data sets are $u_1 = \{t_1, t_2\}$, $u_2=\{t_2, t_3, t_4\}$, and $u_3=\{t_3, t_5\}$. The coalition plan $\mathcal{P}=u_1 \cup u_2 \cup u_3$.  Thus, the coalition data set $D=\{t_1, t_2, t_3, t_4, t_5\}$.  In this example, every synthesis is a single-owner one.
\qed
\end{example}
}
As demonstrated in Example~\ref{ex:assemblage}, for a tuple $t \in D$ in the coalition set, there may exist more than one synthesis, and even a data owner may participate in more than one synthesis of a tuple $t$ in the coalition set.  A synthesis $U=\{u_{i_1}, \ldots, u_{i_m}\}$ is called a \textbf{multi-owner synthesis} if $\|U\| \geq 2$. A synthesis $U$ is a \textbf{minimal synthesis} of tuple $t \in D$ if no proper subset of $U$ is still a synthesis of $t$.

\begin{example}[Syntheses]\rm
Data owners $u_1$ and $u_2$ have schemas $R_1=R_2 = (person1, person2)$ representing which people are mutual friends.  Suppose the data sets $u_1=\{(a, b), (a, c)\}$ and $u_2=\{(b, c)\}$, the target schema $R=(person, person)$ and the coalition plan $\mathcal{P}=u_1 \cup u_2 \cup \mbox{\emph{Proj}}_{R_1.person1, R_2.person2}(u_1 \bowtie_{u_1.person2=u_2.person1} u_2)$.

For tuple $t=(a, c)$ in the coalition set $D$, $\{u_1\}$ is a single-owner synthesis and $\{u_1, u_2\}$ is a multi-owner synthesis. $\{u_1\}$ is a minimal synthesis but $\{u_1, u_2\}$ is not.
\qed
\end{example}

Intuitively, non-minimal syntheses contain redundant data owners.  In a non-minimal synthesis, a data owner not in any minimal synthesis does not contribute to the production of the tuple.  According to the zero element requirement in Shapley fairness, such a data owner should not get any reward in this non-minimal synthesis.  
\nop{
\begin{theorem}[Minimal syntheses]\label{thm:minimal}
Let $D=\{t_1, \ldots, t_l\}$ be a coalition set produced by a coalition by data owners $\mathcal{U}=\{u_1, \ldots, u_n\}$. Under the independent utility assumption, for every data owner $u_i$ $(1 \leq i \leq n)$, the Shapley value of $u_i$ depends on only the minimal syntheses.

\proof 
According to Theorem~\ref{thm:independent}, $\psi(u_i)=\sum_{j=1}^l \psi_{t_j}(u_i)$.  $\psi_{t_j}(u_i)$ can be calculated using Equation~\ref{eq:shapley}.  In Equation~\ref{eq:shapley}, for any subset $\mathcal{S} \subseteq \mathcal{U}$, $Utility_{t_j}(\mathcal{S})$ is either $Utility(t_j)$ or $0$.  $Utility(\mathcal{S})=Utility(t_j)$ if and only if $\mathcal{S}$ contains a minimal synthesis. Therefore, the Shapley value of $u_i$ depends on only the minimal syntheses.
\qed
\end{theorem}

Theorem~\ref{thm:minimal} indicates that, when we calculate Shapley value under independent utility, we only need to consider minimal syntheses.
}
Formally, let $\mathcal{U}_t=\{u_{i_1}, \ldots, u_{i_m}\}$ be the set of data owners each of which participates in at least one minimal synthesis of $t\in D$ in the coalition set.  We call $\mathcal{U}_t$ the \textbf{set of minimal synthesis owners} with respect to $t$.  Clearly, $\mathcal{U}_t \subseteq \mathcal{U}$.  We have the following result.

\begin{theorem}[Minimal syntheses]\label{thm:minimal}
Let $\mathcal{U}=\{u_1, \ldots, u_n\}$ be a set of data owners in coalition.  For a tuple $t \in D$ in the coalition set, let $\mathcal{U}_t=\{u_{i_1}, \ldots, u_{i_m}\}$ be the set of minimal synthesis owners with respect to $t$. Under independent utility, for each owner $u_{i_j}$ $(1 \leq j \leq m)$, the Shapley value
\begin{equation}\label{eq:minimal-syn}
\begin{split}
\psi_t(u_{i_{j}}) & = \frac 1 {\|\mathcal{U}_t\|!} \sum_{\pi' \in \Pi(\mathcal{U}_t)}(Utility_t(P^{\pi'}_{u_{i_j}}\cup\{u_{i_j}\})-Utility_t(P^{\pi'}_{u_{i_j}}))\\
& = \frac{1}{\|\mathcal{U}_t\|} \sum_{\mathcal{S} \subseteq \mathcal{U}_t\setminus\{u_{i_j}\}} \frac{Utility_t(\mathcal{S}\cup \{u_{i_j}\})-Utility_t(\mathcal{S})}{\binom{\|\mathcal{U}_t\|}{\|\mathcal{S}\|}}
\end{split}
\end{equation}
and for any other data owner $u \in \mathcal{U}\setminus \mathcal{U}_t$, $\psi_t(u)=0$.

\proof 
Using Equation~\ref{eq:shapley-permutation}, we have:
$$\psi_t(u_{i_{j}}) = \frac 1 {\|\mathcal{U}\|!} \sum_{\pi \in \Pi(\mathcal{U})}(Utility_t(P^{\pi}_{u_{i_j}}\cup\{u_{i_j}\})-Utility_t(P^{\pi}_{u_{i_j}})) $$

For a permutation $\pi \in \Pi(\mathcal{U})$ and a permutation $\pi' \in \Pi(\mathcal{U}_t)$, $\pi$ is said to \textbf{subsume} $\pi'$, denoted by $\pi' \preceq \pi$, if for every pair of data owners $u_x, u_y$ in $\pi'$ such that $u_x$ precedes $u_y$, $u_x$ also precedes $u_y$ in $\pi$.  In other words, as $\mathcal{U}_t$ is a subset of $\mathcal{U}$, $\pi'$ and $\pi$ are consistent in the order on all data owners in $\mathcal{U}_t$.

For any permutations $\pi \in \Pi(\mathcal{U})$ and $\pi' \in \Pi(\mathcal{U}_t)$ such that $\pi' \preceq \pi$, we show $Utility_t(P^{\pi}_{u_{i_j}}) = Utility_t(P^{\pi'}_{u_{i_j}})$.  This is because $Utility_t(P^{\pi}_{u_{i_j}})=Utility(t)$ if $P^{\pi}_{u_{i_j}}$ contains at least a minimal synthesis of $t$; or otherwise $0$.   Those data owners in the minimal syntheses are all retained in $\pi'$.  

Similarly, we can show $Utility_t(P^{\pi}_{u_{i_j}}\cup\{u_{i_j}\}) = Utility_t(P^{\pi'}_{u_{i_j}}\cup\{u_{i_j}\})$.  

For each permutation $\pi' \in \Pi(\mathcal{U}_t)$, the number of permutations $\pi \in \Pi(\mathcal{U})$ such that $\pi' \preceq \pi$ is 
$$\binom{\|\mathcal{U}\|}{\|\mathcal{U}_t\|}(\|\mathcal{U}\|-\|\mathcal{U}_t\|)!=\frac{\|\mathcal{U}\|!}{\|\mathcal{U}_t\|!}$$
which is a constant.
Thus,
\begin{align*}
\psi_t(u_{i_j}) =& \frac 1 {\|\mathcal{U}\|!} \sum_{\pi \in \Pi(\mathcal{U})}(Utility_t(P^{\pi}_{u_{i_j}}\cup\{u_{i_j}\})-Utility_t(P^{\pi}_{u_{i_j}})) \\
=& \frac 1 {\|\mathcal{U}\|!} \big( \sum_{\pi' \in \Pi(\mathcal{U}_t)}(Utility_t(P^{\pi'}_{u_{i_j}}\cup\{u_{i_j}\}) \\
 & -\, Utility_t(P^{\pi'}_{u_{i_j}}) \big) \frac{\|\mathcal{U}\|!}{\|\mathcal{U}_t\|!}\\
=& \frac 1 {\|\mathcal{U}_t\|!} \sum_{\pi' \in \Pi(\mathcal{U}_t)}(Utility_t(P^{\pi'}_{u_{i_j}}\cup\{u_{i_j}\})-Utility_t(P^{\pi'}_{u_{i_j}}))\\
\end{align*}

Following the equivalence between Equations~\ref{eq:shapley} and~\ref{eq:shapley-permutation}, we have the second form in Equation~\ref{eq:minimal-syn}.

For any other data owners $u$ who are not in the set of minimal synthesis owners, due to the zero element in the Shapley fairness, the Shapley value is $0$.
\qed
\end{theorem}

In the practice of data assemblage, although there may be many data owners, typically there are only very few minimal syntheses of one tuple in the coalition set.  That is, $\|\mathcal{U}_t\|$ is dramatically smaller than $\|\mathcal{U}\|$.  Theorem~\ref{thm:minimal} dramatically reduces the amount of computation for Shapley value even if we want to use a brute-force  approach to compute the exact Shapley value.

In this paper, we assume that syntheses can be obtained in the data assemblage process following the coalition plan.  To find minimal syntheses, for each tuple in the coalition set, we only need to check and remove those syntheses that are proper supersets of some other syntheses.  Since for one tuple in the coalition set, there are only a small number of syntheses, the computation cost is often small.

\subsection{Shapley Value When Only Single-owner Syntheses Exist}\label{sec:single-only}

Let us consider a simple case where a data buyer can simply obtain a tuple $t$ from data owners separately without any interaction among different data owners.  In other words, in this case, all minimal syntheses of $t$ are single-owner syntheses.  Please note that a tuple may be extracted from more than one data owner, and thus the corresponding reward needs to be distributed to all data owners who can contribute the tuple.

How can we calculate the Shapley value for each data owner where there are only single-owner syntheses?  We have the following closed form rule.

\begin{theorem}[Single-owner syntheses only]\label{thm:horizontal}
Let $\mathcal{U}$ be a set of data owners.  For tuple $t \in D$ in the coalition set, if all minimal syntheses of $t$ are single-owner syntheses $\{u_{i_1}\}, \ldots, \{u_{i_m}\}$ $(\{u_{i_1}, \ldots, u_{i_m}\} \subseteq \mathcal{U})$, then, for each data owner $u_{i_j}$ $(1 \leq j \leq m)$, the Shapley value 
\begin{equation}\label{eq:horizontal}
\psi_t(u_{i_j})=\frac {Utility(t)}{m}
\end{equation}
and for any other data owner $u \in \mathcal{U}\setminus \{u_{i_1}, \ldots, u_{i_m}\}$, $\psi_t(u)=0$.

\proof
According to Equation~\ref{eq:shapley-permutation}, for each tuple $t \in D$, the Shapley value of $u$ is 
$$\psi_t(u)=\frac 1 {\|\mathcal{U}\|!} \sum_{\pi \in \Pi(\mathcal{U})}(Utility_t(P^\pi_{u}\cup\{u\})-Utility_t(P^\pi_{u}))$$
For any subset of data owners $\mathcal{X}$, $Utility_t(\mathcal{X})$ can only take two possible values. If there exists at least one data owner $u_{i_j} \in \mathcal{X}$ $(1 \leq j \leq m)$, then $Utility_t(\mathcal{X})=Utility(t)$; otherwise $0$.  Therefore, $Utility_t(P^\pi_{u}\cup\{u\})-Utility_t(P^\pi_{u})$ takes one of the following two cases.

In the first case, $u \in \{u_{i_1}, \ldots, u_{i_m}\}$, that is, $u=u_{i_j}$ for some $1 \leq j \leq m$. $Utility_t(P^\pi_{u_{i_j}}\cup\{u\})-Utility_t(P^\pi_{u_{i_j}})=Utility(t)$ if and only if all data owners in $P^\pi_{u_{i_j}}$ do not form a single-owner synthesis.  Since in total there are $m$ syntheses, the probability that $u_{i_j}$ is before every data owners in $\{u_{i_1}, \ldots, u_{i_m}\} \setminus \{u_{i_j}\}$ in a permutation is $\frac 1 m$, that is,
\begin{equation}\nonumber
\begin{split}
\psi_t(u_{i_j}) &=\frac 1 {\|\mathcal{U}\|!} \sum_{\pi \in \Pi(\mathcal{U})}(Utility_t(P^\pi_{u_{i_j}}\cup\{u_{i_j}\})-Utility_t(P^\pi_{u_{i_j}}))\\
&=\frac {Utility(t)}{m}\\
\end{split}
\end{equation}

In the second case, $u \not\in \{u_{i_1}, \ldots, u_{i_m}\}$. According to Theorem~\ref{thm:minimal}, $Utility_t(P^\pi_{u}\cup\{u\})-Utility_t(P^\pi_{u})=0$.
\qed
\end{theorem}

Note that Equation~\ref{eq:horizontal} can also be proved easily using the balance and symmetry properties of Shapley fairness.  We omit the details here.
The general Shapley value computation is NP-hard.  However, Theorem~\ref{thm:horizontal} shows that, under independent utility, when only single-owner syntheses exist, Shapley value has a closed form solution in constant time. 

\label{sec:unique}

\subsection{Can the Single-owner Solution Be Straightforwardly Generalized?}\label{sec:horizontal-ext}

Equation~\ref{eq:horizontal} gives a simple yet elegant closed form to calculate the Shapley value for an individual data owner when all syntheses are single-owner.  Naturally, one immediate question is whether it can be extended to handle multi-owner syntheses.

For a tuple $t \in D$ in a coalition data set and a data owner $u$, each of the set of minimal synthesis owners contributes to producing some instances of $t$. Can we extend Equation~\ref{eq:horizontal} to 
\begin{equation}\label{eq:vertical-ext1}
\psi_t(u)=\frac {Utility(t)}{\|\mathcal{U}_t\|}
\end{equation}
when there are multi-owner syntheses?  Unfortunately, this does not hold, as shown in the following example.

\begin{example}[Counter examples]\label{ex:horizontal-ext1}\rm
Suppose data owner $u_1=\{(a, b)\}$ in schema $R_1=(A, B)$, data owners $u_2$ and $u_3$ have identical data $u_2=u_3=\{(b, c)\}$ in schema $R_2=R_3=(B, C)$.  The coalition plan $\mathcal{P}=(u_1 \bowtie u_2) \cup (u_1 \bowtie u_3)$, and thus the coalition set $D=\{(a, b, c)\}$.  Let $Utility(a, b, c)=1$.

Using Equations~\ref{eq:shapley} or~\ref{eq:shapley-permutation}, we can easily calculate the Shapley value of the data owners, $\psi(u_1)=\frac 2 3$ and $\psi(u_2)=\psi(u_3)=\frac 1 6$.

Using Equation~\ref{eq:vertical-ext1}, we have $\|\mathcal{U}_t\|=3$, since $\{u_1, u_2\}$ and $\{u_1, u_3\}$ are two minimal syntheses of $(a, b, c)$.  Therefore, Equation~\ref{eq:vertical-ext1} calculates the same value, $\frac 1 3$ for each of $u_1$, $u_2$ and $u_3$, which are not equal to their Shapley values.
\qed
\end{example}

Example~\ref{ex:horizontal-ext1} indicates that, when multiple owners need to work together to form a tuple in the coalition set, an owner who has more ways to collaborate with other owners can claim a higher Shapley value.  In other words, when there are multi-owner syntheses, not every data owner contributing to a tuple in the coalition set gains the same for the Shapley value.

Example~\ref{ex:horizontal-ext1} also suggests that the ways data owners assemble a tuple in the coalition set matter.  Can we equivalently split the utility of a tuple in the coalition set among all the multi-owner syntheses and then, within each synthesis, equivalently split the utility among all data owners participating?

\begin{example}[Example~\ref{ex:horizontal-ext1} continued]\label{ex:horizontal-ext2}\rm
In the case in Example~\ref{ex:horizontal-ext1}, there are $2$ minimal multi-owner syntheses of $(a, b, c)$ in the coalition set, $U_1=\{u_1, u_2\}$ and $U_2=\{u_1, u_3\}$. One may think, we may evenly split the utility $1$ of the tuple $(a, b, c)$ in the coalition set between the $2$ syntheses,  thus each synthesis receives utility $\frac 1 2$.  Then, each data owner obtains an even share in each synthesis between all the owners participating.  For example, $u_1$ obtains half of the utility $\frac 1 2$ in the synthesis $U_1$ where $u_1$ and $u_2$ join their tuples to form $(a, b, c)$.  The utility allocated to $u_1$ is $\frac 1 2 \times \frac 1 2 + \frac 1 2 \times \frac 1 2 = \frac 1 2$, it still does not match $\psi(u_1)=\frac 2 3$.
\qed
\end{example}

The failures of the above attempts clearly show that computing Shapley value where there are multi-owner syntheses is far from trivial.  Some straightforward extensions to Equation~\ref{eq:horizontal}, the closed form solution to the single-owner only situation, cannot capture Shapley value in a general scenario correctly.

\subsection{Unique Multi-owner Synthesis and Shapley Value Closed Form}\label{sec:synthesis}

Now, let us consider a still specific situation involving multi-owner syntheses.  It is more general than the case discussed in Section~\ref{sec:single-only} where there are only single-owner syntheses.

For a tuple $t \in D$ in a coalition set, if there exists only one minimal multi-owner synthesis $U$, then $U$ is called the \textbf{unique multi-owner synthesis} of $t$.  Unique multi-owner synthesis enables a closed form calculation of Shapley value.

\begin{theorem}\label{thm:unique-multi}
For a tuple $t$ in a coalition set, if there is a unique multi-owner synthesis $U=\{u_{i_1}, \ldots, u_{i_m}\}$ $(m \geq 2)$ and $k$ single-owner syntheses of $t$, then the Shapley value of each data owner $u_{i_j} \in U$ $(1 \leq j \leq m)$ is
$$\psi_{t}(u_{i_j})=\frac{Utility(t)}{(m+k)\binom{m+k-1}{m-1}}$$
and the Shapley value of each data owner $u$ contributing to a single-owner synthesis is 
$$\psi_t(u)=\frac{Utility(t)}k\big(1-\frac{m}{(m+k)\binom{m+k-1}{m-1}}\big).$$

\proof
Let $\mathcal{U}$ be the set of all data owners.  Let $\{u_{l_1}\}$, \ldots, $\{u_{l_k}\}$ be the $k$ single-owner syntheses.  Apparently, $u_{l_x} \not\in U$ for $1 \leq x \leq k$, otherwise, $U$ is not minimal.  Therefore, $\mathcal{U}_t=\{u_{i_1}$, $\ldots$, $u_{i_m}$, $u_{l_1}$, $\ldots$, $u_{l_k}\}$.

According to Theorem~\ref{thm:minimal}, we have 
$$\psi_t(u_{i_j}) = \frac 1 {\|\mathcal{U}_t\|} \sum_{\mathcal{S} \subseteq \mathcal{U}_t \setminus \{u_{i_j}\}} \frac{Utility_t(\mathcal{S} \cup (u_{i_j}))- Utility_t(\mathcal{S})}{\binom{\|\mathcal{U}_t\|-1} {\|\mathcal{S}\|}}$$
Obviously, since there is only one unique multi-owner synthesis $U$ and $k$ single-owner syntheses,  $Utility_t(\mathcal{S} \cup (u_{i_j}))- Utility_t(\mathcal{S})=Utility(t)$ if and only if $S = U \setminus \ u_{i_j}$.  In all other situations, $Utility_t(\mathcal{S}$ $\cup (u_{i_j}))- Utility_t(\mathcal{S})=0$. 
In other words, in the only non-zero case, $\|\mathcal{U}_t\| = m+k$ and $\|\mathcal{S}\| = m-1$.  Thus, we have
$$\psi_{t}(u_{i_j})=\frac{Utility(t)}{(m+k)\binom{m+k-1}{m-1}}$$ 

Now, let us consider the data owners in the single-owner syntheses.  According to the balance requirement in Shapley fairness, 
$$\sum_{x=1}^{k} \psi_t(u_{l_x}) = Utility(t) - \sum_{j=1}^{m} \psi_{t}(u_{i_j})$$
According to symmetry requirement in Shapley fairness, for each data owner $u \in \{u_{l_1}, ...,u_{l_k} \}$, 
\begin{align*}
\psi(u) &= \frac{1}{k} \sum_{x=1}^{k} \psi_t(u_{l_x}) 
= \frac{Utility(t) - \sum_{j=1}^{m} \psi_{t}(u_{i_j})}{k} \\
&=\frac{Utility(t)}k\big(1-\frac{m}{(m+k)\binom{m+k-1}{m-1}}\big). 
\end{align*}
\qed
\end{theorem}

In Theorem~\ref{thm:unique-multi}, if there is no unique multi-owner synthesis at all, and there are only single-owner syntheses, then the Shapley value of a data owner who contributes to a single-owner synthesis is the same as computed in Theorem~\ref{thm:horizontal}.  Thus, Theorem~\ref{thm:unique-multi} is a general result covering Theorem~\ref{thm:horizontal}. 

Under independent utility, when there exist only one unique multi-owner synthesis of size $m$ and optionally $k$ single-owner syntheses, using Theorem~\ref{thm:unique-multi}, we have a closed form solution for Shapley value in linear time $O(\min(m, k))$.

\section{Computing Shapley Value Under Independent Utility}
\label{sec:algo}

In this section, we develop two algorithms for the general situation beyond the two special cases discussed in Sections~\ref{sec:single-only} and~\ref{sec:synthesis} and present the overall Shapley value computation method integrating all possible cases.  

\subsection{The Synthesis-combination (SC) Algorithm}

A straightforward approach to compute, for each tuple $t$ in a coalition set, the Shapley value of data owners $\psi_t(u)$ is to apply Equation~\ref{eq:shapley-permutation}.  The equation requires us to compute the utility values $Utility_t(P^\pi_u)$ and $Utility_t(P^\pi_u \cup \{u\})$ for every possible permutation $\pi$ of users. A na\"ive method is to check whether the data owners in $P^\pi_u$ and $P^\pi_u \cup \{u\}$ can synthesize $t$ according to the coalition plan.  

To speed up the computation and reduce the number of permutations needed to consider, we propose a  \textbf{synthesis-combination (SC) algorithm}.  The general idea is that we materialize all syntheses of every tuple in the coalition set as a byproduct of computing the coalition set.  Then, when we apply Equation~\ref{eq:shapley-permutation}, we use the combinations of the syntheses to cover all permutations and obtain the utility values instead of computing the utility from scratch again and again.

As the first step, we compute the coalition set $D$ according to the coalition plan $\mathcal{P}$.  In this process, for each tuple $t \in D$, we record the set of minimal syntheses $t.S$.

\begin{example}[Computing Syntheses]\label{ex:compute-syntheses}\rm
Consider the data owners and their data as well as the coalition plan in Example~\ref{ex:horizontal-ext1}.  Following the coalition plan, we have the coalition set $D=\{(a, b, c)\}$.  There are two multi-owner syntheses of tuple $t=(a, b, c)$.  Thus, $t.S=\{\{u_1, u_2\}, \{u_1, u_3\}\}$.
\qed
\end{example}

To evaluate Equation~\ref{eq:shapley-permutation} efficiently, we only need to find the permutations $\pi$ such that $Utility_t(P^\pi_u \cup \{u\}) - Utility_t(P^\pi_u)=Utility(t)$.  
Assume there are in total $m_u$ minimal syntheses $U_{i_1}$, $\ldots$, $ U_{i_{m_u}}$ of $t$ that contain $u$, and $m_{\bar{u}}$ minimal syntheses $U_{j_1}, \ldots, U_{j_{m_{\bar{u}}}}$ that do not contain $u$. There are two situations where $Utility_t(P^\pi_u \cup \{u\}) - Utility_t(P^\pi_u)=0$. First, if $P^\pi_u \cup \{u\}$ does not contain any minimal synthesis in $\{U_{i_1},\ldots,U_{i_m}\}$, $Utility_t(P^\pi_u \cup \{u\}) - Utility_t(P^\pi_u)=0$. Second, when $P^\pi_u \cup \{u\}$ contains at least one minimal synthesis in $\{U_{i_1},\ldots,U_{i_m}\}$, if $P^\pi_u \cup \{u\}$ contains at least one minimal synthesis in $\{U_{j_1}, \ldots, U_{j_{m_{\bar{u}}}}\}$, $Utility_t(P^\pi_u \cup \{u\}) - Utility_t(P^\pi_u)=0$.  Thus, in order to have $Utility_t(P^\pi_u \cup \{u\}) - Utility_t(P^\pi_u)=Utility(t)$, $P^\pi_u \cup \{u\}$ must contain at least one minimal synthesis in $\{U_{i_1},\ldots,U_{i_m}\}$ and does not contain any minimal synthesis in $\{U_{j_1}, \ldots,  U_{j_{m_{\bar{u}}}}\}$ .

To formalize the above insight, let $$V_1=\{\pi \in \Pi(\mathcal{U})|\exists 1 \leq k \leq m_u: U_{i_k} \subseteq P_u^\pi \cup \{u\} \}$$
and
$$V_2=\{\pi \in V_1|\exists 1 \leq k \leq m_{\bar{u}}: U_{j_k} \subseteq P_u^\pi\}$$
Then, Equation~\ref{eq:shapley-permutation} can be rewritten to $\psi_t(u)=  Utility(t) \cdot \frac {\|V_1\|-\|V_2\|} {\|\mathcal{U}\|!}$.

Now, let us work on how to calculate $V_1$ and $V_2$.  For a synthesis $U$ of $t$ and a data owner $u$ such that $u \in U$,  denote by $V_{U\succ u}$ the set of permutations where $U \setminus \{u\}$ precedes $u$. We have the following nice properties.

\begin{lemma}\label{lem:lemma}
Given a set of data owners $\mathcal U$, for a synthesis $U$ of $t$ and a data owner $u \in U$, $\frac{\|V_{U \succ u}\|}{\|\mathcal{U}\|!} = \frac 1 {\|U\|}$.  Moreover, if there are $m$ syntheses $U_1, \ldots, U_m$ such that $u \in U_i$ $(1 \leq i \leq m)$, then 

$$\bigcap_{i=1}^m V_{U_i \succ u} = V_{(\cup_{i=1}^m U_i)\succ u}$$
Last, 
\begin{equation}\label{eq:precede-union}
\|\bigcup_{i=1}^m V_{U_i \succ u}\| =  \sum_{i=1}^m (-1)^{i+1}\Big(\sum_{1\leq j_1< \cdots < j_i \leq m}\|V_{(\cup_{k=1}^i U_{j_k}) \succ u}\|\Big)
\nop{
\begin{split}
\|\bigcup_{i=1}^m V_{U_i \succ u}\| = & \|V_{(\cup_{i=1}^m U_i) \succ u}\| - \sum_{1 \leq i < j \leq n} \|V_{U_i \cup U_j \succ u}\| \\
& + \sum_{1 \leq i < j < k \leq m} \|V_{U_i \cup U_j \cup U_k \succ u}\| + \cdots \\
& + (-1)^{m-1} \|V_{(\cup_{i=1}^m U_i)\succ u}\| \\
\end{split}
}
\end{equation}

\proof
Since $u \in U$, $\|V_{U\succ u}\|=\binom{\|\mathcal{U}\|}{\|U\|}(\|U\|-1)!(\|\mathcal{U}\|-\|U\|)!$. Therefore, 
\begin{equation}\nonumber
\begin{split}
\frac{\|V_{U \succ u}\|}{\|\mathcal{U}\|!}
&= \frac{\binom{\|\mathcal{U}\|}{\|U\|}(\|U\|-1)!(\|\mathcal{U}\|-\|U\|)!}{\|\mathcal{U}\|!}\\
&=\frac{\frac{\|\mathcal{U}\|!}{(\|\mathcal{U}\|-\|U\|)!\|U\|!}({\|U\|-1)!}(\|\mathcal{U}\|-\|U\|)!}{\|\mathcal{U}\|!}= \frac{1}{\|U\|}  \\
\end{split}
\end{equation}

The second property holds because, according to the definition, the set $\bigcap_{i=1}^m V_{U_i \succ u} $ contains all permutations $\pi$ where all data owners in $U_1\setminus \{u\}, \ldots, U_m \setminus \{u\}$ all precede $u$, that is, $V_{(\cup_{i=1}^m U_i)  \succ u}$.  

Last, $\|\bigcup_{i=1}^m V_{U_i \succ u}\|$ can be calculated using the set union cardinality formula and the second property in this theorem.
\qed
\end{lemma}

Now, for a tuple $t\in D$ in a coalition set, we are ready to compute $V_1$ and $V_2$ with respect to $t$ and give a formula to calculate $\psi_t(u)$ for each data owner $u$.

\begin{theorem}\label{thm:syn-comb}
For a data owner $u$ and a tuple $t$ in a coalition set $D$, let $W_u=\{U_{i_1}, \ldots, U_{i_{m_u}}\}$ be the set of minimal syntheses that contain $u$, $W_{\bar{u}}=\{U_{j_1}, \ldots, U_{j_{m_{\bar{u}}}}\}$ be the set of minimal syntheses that do not contain $u$.
Let
$$\nu(W_u)=\sum_{\substack{X\subseteq W_u\\ \|X\| \geq 1}} \frac{(-1)^{\|X\|+1}}{\| \cup_{U_x \in X} U_x\|}$$ and 
$$\tau(W_u, W_{\bar{u}})=\sum_{\substack{X \subseteq W_u \times W_{\bar{u}}\\ \|X\| \geq 1}}\frac{(-1)^{\|X\|+1}}
{\|\cup_{(U_x, U_y) \in X}(U_x \cup U_y)\|}.$$ Then,
\begin{equation}\label{eq:syn-comb}
\psi_t(u)=Utility(t)(\nu(W_u)-\tau(W_u, W_{\bar{u}})).
\end{equation}

\proof
Following the definitions, we have $V_1 = \cup_{x=1}^{m_u} V_{ U_{i_x} \succ u}$ and $V_2 = \bigcup_{\substack{U_x \in W_u\\ U_y \in W_{\bar{u}}}} V_{(U_x \cup U_y) \succ u}$. We calculate $ \psi_t(u)$ as follows.
    \begin{equation}\label{eq:tmp1}
    \begin{split}
        \psi_t(u) 
        =&  Utility(t)  \frac{\|V_1\| - \|V_2\|}{{\|\mathcal{U}\|!}}  \\
        =& Utility(t)  \frac{\| \bigcup_{x=1}^{m_u} V_{ U_{i_x} \succ u} \| - \|  \bigcup_{\substack{U_x \in W_u\\ U_y \in W_{\bar{u}}}} V_{(U_x \cup U_y) \succ u} \|}{\|\mathcal{U}\|!}
\end{split}
\end{equation} 

Applying Equation~\ref{eq:precede-union}, we have 
\begin{equation}\label{eq:tmp2}
\frac{\| \bigcup_{x=1}^{m_u} V_{ U_{i_x} \succ u} \|}{\|\mathcal{U}\|!}= \sum_{x=1}^{m_u}(-1)^{x+1}\Big( \sum_{1 \leq y_1 < \cdots < y_x \leq m_u} \frac {\| V_{(\cup_{z=1}^x U_{i_{y_z}})\succ u}\|}{\|\mathcal{U}\|!}\Big)
\end{equation}

According to Lemma~\ref{lem:lemma}, $\frac{\|V_{U \succ u}\|}{\|\mathcal{U}\|!} = \frac 1 {\|U\|}$, thus, in Equation~\ref{eq:tmp2}, $\frac {\| V_{(\cup_{z=1}^x U_{i_{y_z}})\succ u}\|}{\|\mathcal{U}\|!}=\frac 1 {\| \cup_{z=1}^x U_{i_{y_z}}\|}$.  Thus, Equation~\ref{eq:tmp2} can be further written as
\begin{equation}\label{eq:tmp3}
\begin{split}
\frac{\| \bigcup_{x=1}^{m_u} V_{ U_{i_x} \succ u} \|}{\|\mathcal{U}\|!}=& \sum_{x=1}^{m_u}(-1)^{x+1}\Big( \sum_{1 \leq y_1 < \cdots < y_x \leq m_u} \frac 1 {\| \cup_{z=1}^x U_{i_{y_z}}\|}
\Big)\\
=& \sum_{x=1}^{m_u} \sum_{1 \leq y_1 < \cdots < y_x \leq m_u} \frac{(-1)^{x+1}}{\| \cup_{z=1}^x U_{i_{y_z}}\|}\\
=& \sum_{\substack{X\subseteq W_u\\ \|X\| \geq 1}} \frac{(-1)^{\|X\|+1}}{\| \cup_{U_x \in X} U_x\|} = \nu(W_u)
\end{split}
\end{equation}

Similarly, we can show
\begin{equation}\label{eq:tmp4}
\frac{\|  \bigcup_{\substack{U_x \in W_u\\ U_y \in W_{\bar{u}}}} V_{(U_x \cup U_y) \succ u} \|}{\|\mathcal{U}\|!} = \tau(W_u, W_{\bar{u}})
\end{equation}

Plugging Equations~\ref{eq:tmp3} and~\ref{eq:tmp4} into Equation~\ref{eq:tmp1}, we establish Equation~\ref{eq:syn-comb} immediately.
\qed
\end{theorem}

The synthesis-combination method reduces the number of permutations that need to be computed. The complexity of the algorithm is independent from the number of data owners and, instead, is exponential to $\max(m_u, m_u m_{\bar{u}})$. Assuming the cost for each set union and arithmetic operations being bounded by a constant, the time complexity of the algorithm is $O(2^{\max(m_u, m_u m_{\bar{u}})})$.
It is very fast when there are only very few minimal syntheses with respect to a tuple in the coalition set, although there may be many data owners.  
However, if there are many minimal syntheses on a tuple in the coalition set, the algorithm is costly %
since $\max(m_u, m_u m_{\bar{u}})$ is large in such a case.  In the worst case, $\max(m_u, m_u m_{\bar{u}})$  may even exceed the number of data owners.

\subsection{The Synthesis-look-up (SL) Algorithm}

When there are many minimal syntheses with respect to a tuple in a coalition set, how can we compute the Shapley value fast?  We propose a \textbf{synthesis-look-up (SL) algorithm}.  The general idea is that we still materialize all syntheses of every tuple in the coalition set as a byproduct of computing the coalition set and use the second form in Equation~\ref{eq:minimal-syn} to calculate the Shapley value, that is,
$$\psi_t(u) = \frac 1 {\|\mathcal{U}_t\|} \sum_{\mathcal{S} \subseteq \mathcal{U}_t \setminus \{u\}} \frac{Utility_t(\mathcal{S} \cup (u))- Utility_t(\mathcal{S})}{\binom{\|\mathcal{U}_t\|-1} {\|\mathcal{S}\|}}$$

For each subset $\mathcal{S} \subseteq \mathcal{U}_t\setminus \{u\}$ in Equation~\ref{eq:shapley}, instead of computing the utility from scratch, we obtain the utility by checking whether $\mathcal{S}$ and $\mathcal{S} \cup \{u\}$ contain a synthesis.

Specifically, to calculate the Shapley value $\psi_t(u)$ of data owner $u$ with respect to tuple $t \in D$ in the coalition set, we divide all minimal syntheses of $t$ into two sets.  Let $W_u=\{U_{i_1}, \ldots, U_{i_{m_u}}\}$ be the set of minimal syntheses that contain $u$, and $W_{\bar{u}}=\{U_{j_1}, \ldots, U_{j_{m_{\bar{u}}}}\}$ be the set of minimal syntheses that do not contain $u$. Clearly, for each subset $\mathcal{S} \subseteq \mathcal{U}_t\setminus \{u\}$ in Equation~\ref{eq:shapley}, $Utility_t(\mathcal{S}\cup\{u\})-Utility_t(\mathcal{S})=Utility(t)$ if and only if there exists a $k_u$ $(1 \leq k_u \leq m_u)$ such that $U_{i_{k_u}} \subseteq \mathcal{S}$ and there does not exist any $k_{\bar{u}}$ $(1 \leq k_{\bar{u}} \leq m_{\bar{u}})$ such that $U_{i_{k_{\bar{u}}}} \subseteq \mathcal{S}$.  This condition can be checked using the materialized minimal syntheses.

The complexity of the synthesis-look-up algorithm does not rely on the number of minimal syntheses.  Instead, it is exponential to the number of data owners participating in the minimal syntheses, that is, the size of the set of minimal synthesis owners $\| \mathcal{U}_t \|$. Assuming the cost for each super set checking and arithmetic operations being bounded by a constant, the time complexity of the algorithm is $O(2^{\|\mathcal{U}_t \|})$. When there are not many data owners participating in the minimal syntheses, the algorithm is fast, though there may still be many minimal syntheses.

\begin{algorithm}[t]
\KwInput{tables $D_1, \ldots, D_n$ from data owners $u_1, \ldots, u_n$, coalition plan $\mathcal{P}$, hyper-parameter $\gamma$}
\KwOutput{coalition set $D$ and for each data owner $u_i$ $(1\leq i \leq n)$, the Shapley value $\psi(u_i)$}
compute the coalition set $D$ according to the coalition plan $\mathcal{P}$, for each tuple $t \in D$, record the set of minimal syntheses $t.S$;

\ForEach{tuple $t \in D$}{
	\uIf {$t$ has only single-user syntheses}
		{apply Theorem~\ref{thm:horizontal}\;}
	\uElseIf {$t$ has unique multi-owner synthesis}
	{apply Theorem~\ref{thm:unique-multi}\;}
	\Else{
		  \ForEach{data owner $u \in \mathcal{U}_t$} {
	    	\uIf {$\| \mathcal{U}_t \| > \gamma \cdot \ \max(m_u, m_u m_{\bar{u}})$}
			    {apply the SC algorithm (Theorem~\ref{thm:syn-comb})\;}
		    \Else{apply the SL algorithm\;}
	    }
	}
}
\caption{IUSV: computing Shapely value under independent utility.}\label{algo:main}
\end{algorithm}

\subsection{Summary: the Independent Utility Shapley Value (IUSV) Approach}

The synthesis-combination method and the synthesis-look-up \allowbreak method have their individual advantages and complement to each other.  Moreover, we have the special cases when a unique multi-owner synthesis exists or there are only single-owner syntheses. \Cref{algo:main} presents our independent utility Shapley value (IUSV) approach.
In order to coordinate the synthesis-combination algorithm and the synthesis-look-up algorithm to handle different situations, the IUSV algorithm uses a hyper-parameter $\gamma$.  When there are relatively fewer minimal syntheses, the synthesis-combination method is used, otherwise, the synthesis-look-up method is employed.  We will experimentally examine the effect of the hyper-parameter in Section~\ref{sec:exp}.
The overall time complexity of the algorithm is $O(\sum_{t \in D} \min\{2^{\|\mathcal{U}_t \|}, 2^{\max(m_u, m_u m_{\bar{u}})}\}).$ %

\section{Performance Evaluation}\label{sec:exp}

In this section, we empirically evaluate the performance of our proposed IUSV approach and compare with two representative baselines using two benchmark real data sets. We first describe the experiment setup and then present the experimental results.

\subsection{Experiment Setup}

We compare our method with two baselines as follows.
\begin{itemize}
    \item The \textbf{traditional method (Trad)} computes the exact Shapley value based on Equation~\ref{eq:shapley}.
    \item The \textbf{permutation-based sampling method (Perm)} approximates Shapley value using the Monte-Carlo method according to Equation~\ref{eq:shapley-permutation}~\cite{maleki2013bounding}. We write Perm-$x$ (e.g., Perm-16 and Perm-32) when the permutation-based sampling method takes $x$ permutations as the sample.
\end{itemize}

We implement all of these methods using the Rust programming language~\cite{matsakis2014rust}.
The source codes of our implementation are available at \url{https://github.com/IDEAL-Lab/shapley-value-independent-utility}.

We use two real data sets in our experiments, namely World\footnote{\url{https://dev.mysql.com/doc/world-setup/en/}, accessed on July 1, 2022.} %
and TPC-H\footnote{\url{http://www.tpc.org/tpch/}, accessed on July 1, 2022.}. The World data set consists of $3$ tables with $239$, $984$, and $\fnum{4079}$ records, respectively.
The TPC-H data set has $8$ tables with $5$, $25$, $\fnum{10000}$, $\fnum{150000}$,  $\fnum{200000}$, $\fnum{800000}$, $\fnum{1500000}$, and $\fnum{6001215}$ records, respectively.  The World data set is used to evaluate the performance of the baselines in some settings where they cannot complete on the large TPC-H data set within reasonable time.

We randomly assign records in a data set to data owners in three steps. In the first step, we assign data owners to each table in the data set. Two different scenarios are implemented. 
\begin{itemize}
\item \textbf{EO} (for equal number of owners) assigns $k$ data owners to each table and keeps the number of data owners for each table equal. The only exception is the two smallest tables in the TPC-H data set, where we only assign a single data owner to those two tables since those two tables only have very few records each, $5$ and $25$, respectively.  

\item \textbf{UO} (for unequal number of owners) assigns $k$ data owners to the largest table in a data set and only $2$ data owners to each of the other tables. In this setting, we test how the splitting of tuples in a large table may affect the Shapley value computation.
\end{itemize}

In the second step, after assigning the data owners to each table, we duplicate each record in the original data set several times. The numbers of copies of tuples follow the Zipfian distribution~\cite{newman2005power} with parameter $\alpha$. We also enforce a hard constraint on the maximum number of copies that a tuple can be duplicated, that is, a record cannot have more than $m$ copies.

As the last step, we assign records to data owners. Two scenarios are considered.
\begin{itemize}

\item \textbf{EA} (for equal chance assignment) assigns records to data owners with uniform distribution. That is, if a record has $l$ copies and there are $k$ data owners assigned to the table, then each data owner has a probability $\frac l k$ to get a copy of the record. We constrain that each data owner can only have up to one copy of a record. Within a table, the expected number of records held by each data owner is identical.

\item \textbf{UA} (for unequal chance assignment) assigns records to data owners following the Zipfian distribution, that is, the probability of a data owner obtaining a record obeys the Zipfian distribution with parameter $\beta$. Again, we constrain that each data owner can only have up to one copy of a record. In this case, a small number of data owners hold most of the records in a table. 

\end{itemize}

In total, we have four different settings in setting number of data owners in tables and assigning records to data owners, namely \textsf{UO-UA}, \textsf{EO-UA}, \textsf{UO-EA}, and \textsf{EO-EA}. Through those settings we can observe how data owner distribution may affect the performance of Shapley value computation. By setting \textsf{UO} versus \textsf{EO}, we can observe the effect of the number of data owners in each table. By setting \textsf{UA} versus \textsf{EA}, we can observe the effect of the numbers of records held by data owners. The default parameters used in the experiments are shown in~\cref{tab:experimental-parameters}.

\begin{table}[t]
  \caption{Default System Parameters}%
  \label{tab:experimental-parameters}
  \centering
  \small
  \begin{tabular}{lc}
      \toprule
      \textbf{Parameters} & \textbf{Default value} \\
      \midrule
      Number of data owners $k$ per table & 5 (World), 10 (TPC-H) \\
      Zipfan parameter $\alpha$ & 4.0 \\
      Max copy of tuples $m$ & 3 \\ 
      Zipfan parameter $\beta$ (used in \textsf{UA}) & 3.0 \\
      Hyper-parameter $\gamma$ in Algorithm~\ref{algo:main} & 1.0 \\
      \bottomrule
  \end{tabular}
\end{table}

To evaluate the performance, a coalition plan executes equi-join queries among all tables in a data set.  We assume that the utility of each tuple in a coalition set is $1$.

We use a commodity server with Intel Xeon 2.00GHz E7-4730 CPU and 125GB RAM, running Ubuntu 20.04 LTS to run the experiments. 
We focus on two metrics as follows.
\begin{itemize}

\item \textbf{Runtime} measures the total clock time of computing the Shapley values for all data owners.

\item \textbf{Error rate} evaluates the quality of the approximated Shapley value by the permutation-based sampling method and computes the percentage of miscalculated Shapely value, that is, $error =  \frac{\sum_{u \in \mathcal{U}} | \psi(u) - \widehat{\psi(u)} |}{ \sum_{u \in \mathcal{U}} \psi(u)}$, where $\psi(u)$ and $\widehat{\psi(u)}$ are the exact Shapley value and the approximate Shapley value by the permutation-based sampling method, respectively.  The error rate reflects the ratio of the total accumulated absolute errors against the total Shapley value of all data owners.
\end{itemize}

For our proposed IUSV method, we measure two additional metrics as follows.
\begin{itemize}
    \item \textbf{UMOS rate}  (for unique multi-owner synthesis rate) is the ratio of the number of tuples with only one unique multi-owner synthesis against the number of tuples in the coalition set. Recall that the tuples with only one unique multi-owner synthesis can use \cref{thm:horizontal,thm:unique-multi} to compute Shapley value directly.  Thus, UMOS rate shows the percentage of cases where \cref{thm:horizontal,thm:unique-multi} can be applied.
    \item \textbf{SC rate and SL rate} are, among the total number of times the SC and SL algorithms are called when the multi-owner syntheses of the tuples in the coalition set are not unique, the percentages of the calls to the SC  and SL algorithms, respectively.  These metrics show the relative frequencies the two algorithms are used. 
\end{itemize}

In our experiments, we enforce a timeout of $\fnum{7200}$ seconds, that is, we terminate a program if the runtime exceeds the allowed timeout. %

\begin{table}[t]
  \centering
  \caption{Permutation Method Error Rate Under Default Parameters}\label{tab:error}
    \begin{threeparttable}[b]
      \small%
      \begin{tabular}{ccccc}
        \toprule
        \multirow{2}{*}[-0.5\dimexpr\aboverulesep+\belowrulesep+\cmidrulewidth]{\textbf{Setting}} &
        \multicolumn{2}{c}{\textbf{World}} &
        \multicolumn{2}{c}{\textbf{TPC-H}} \\
        \cmidrule(r){2-3}
        \cmidrule(l){4-5}
        & Perm-16 & Perm-32 & Perm-16 & Perm-32\\
        \midrule
        \textsf{UO-UA} & 29\% & 22\% &  63\% & 37\% \\
        \textsf{EO-UA} & 33\% & 24\% &  58\% & 42\% \\
        \textsf{UO-EA} & 23\% & 18\% &  54\% & 36\% \\
        \textsf{EO-EA} & 23\% & 17\% &  49\% & 37\% \\
        \bottomrule
      \end{tabular}
    \end{threeparttable}
\end{table}

\subsection{Scalability on Number of Data Owners}

\begin{figure*}[t]%
    \centering
    \includegraphics[width=\linewidth]{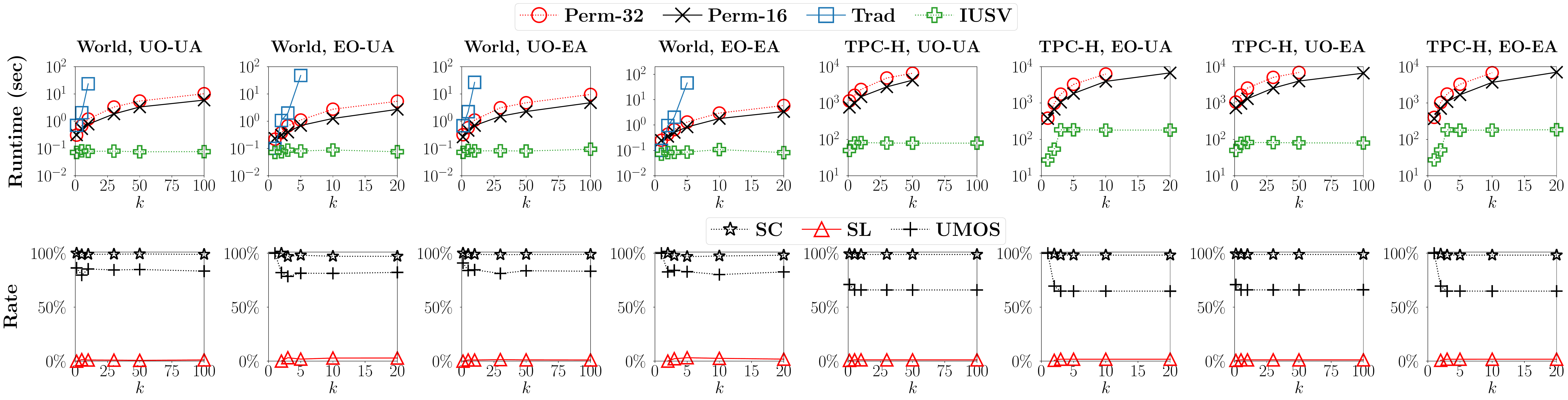}
    \caption{Effect of Number of Data Owners $k$ per Table.}%
    \label{exp-fig:time_vs_owner}%
\end{figure*}

\begin{figure*}[t]%
    \centering
    \includegraphics[width=\linewidth]{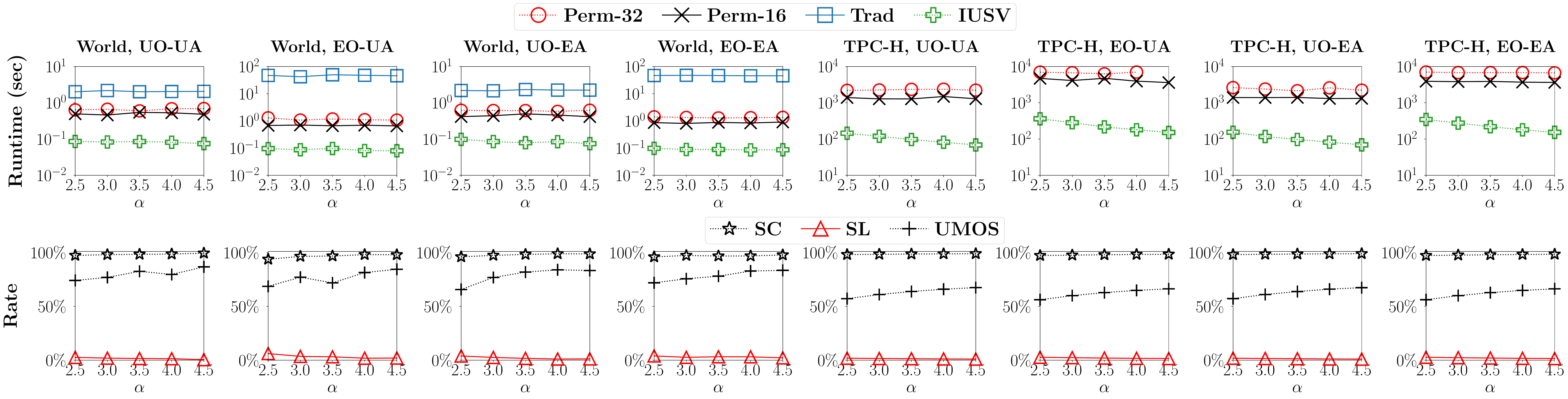}
    \caption{Effect of Zipfian Parameter $\alpha$.}%
    \label{exp-fig:time_vs_alpha}%
\end{figure*}
\Cref{exp-fig:time_vs_owner} shows the performance in the aforementioned four settings when the number of data owners $k$ per table varies. Compared with the traditional method on the World data set, our proposed IUSV method reduces the total runtime by 3 orders of magnitude before the traditional method fails to finish the computation before the timeout. Our IUSV method can efficiently compute the exact Shapely value in all cases.

For the permutation-based sampling method, we choose a small sample size, $16$ or $32$ for the data sets World and TPC-H, since a larger sample size causes the permutation method timeout. To obtain a permutation sample, the method has to compute the equi-join of all tables once and thus is very costly.  With such small samples, the accuracy of the approximation is low. \Cref{tab:error} shows the error rates under the default parameters. As shown, the error rate ranges from $17\%$ to $63\%$.  The errors are significant.  At the same time, even we choose only 16 samples and tolerate a large error (up to $63\%$), Perm-16 is still slower than IUSV, as shown in \Cref{exp-fig:time_vs_owner}.

~\cref{exp-fig:time_vs_owner} shows that the UMOS rate decreases as the number of data owners increases from $1$ or $2$ to a small number (less than $5$ in our experiments) in all settings, and then becomes stable. When the number of data owners increases from $1$ or $2$ to a small number, substantially more records in the coalition set have more than one multi-owner synthesis.  However, when the number of data owners keeps increasing, due to the data sparsity, the UMOS rate becomes stable. The results show that a substantial portion of cases can be handled by~\cref{thm:horizontal,thm:unique-multi} efficiently, which contributes to the efficiency of IUSV significantly.  

The SC and SL rates show that in most of the cases where multi-owner syntheses are not unique, the SC algorithm is used.  At the same time, the SL algorithm handles a small portion of cases that are very costly if the SC algorithm is used.  
The complement between the two methods provides an overall fast solution.

\subsection{Effect of Data Owner and Record Assignment Distributions}

\cref{exp-fig:time_vs_alpha} shows the performance in the aforementioned four settings with respect to the Zipfian parameter $\alpha$, which controls the number of copies of records in tables. In all settings, the runtime of all methods on the World data set is insensitive to $\alpha$ mainly due to the small size of the data set. The runtime of IUSV on TPC-H decreases as $\alpha$ increases mainly for two reasons. First, a larger $\alpha$ assigns more records to one data owner. Therefore, the UMOS rate increases and thus IUSV can apply Theorem~\ref{thm:unique-multi} to calculate Shapley values in more cases. Second, more records held by one data owner leads to decrease in both the number of minimal syntheses and the number of minimal synthesis owners. Consequently, the execution of the SC algorithm becomes faster when the number of minimal synthesis is smaller, and the SL algorithm is faster when the number of minimal synthesis owners is smaller. 

The maximum number of copies of a data record as a hard constraint also has a mild effect similar to the situation where $\alpha$ is small.  The smaller the maximum number, the more data records are held by a data owner.  Limited by space, we omit the details here.

\nop{
\Cref{exp-fig:time_vs_m} shows runtime under the aforementioned four settings when max copy of tuples parameter $m$ is varied.
Similarly we observe $m$ has no impact on the runtime of all methods on the World data set and baseline methods on TPC-H data set. However, we can see that the runtime of IUSV method in TPC-H data set increases when the max copy of tuples $m$ increases. Because when $m$ increases, more data owners may hold the same copy of a record, which leads to a decrease of tuples with  unique multi owner syntheses and an increase of number of minimal syntheses. These factors in turn lead to a longer time in executing SC or SL algorithms. In the end, the runtime of our proposed method increases.
}

\begin{figure}[t]%
    \centering
    \includegraphics[width=\linewidth]{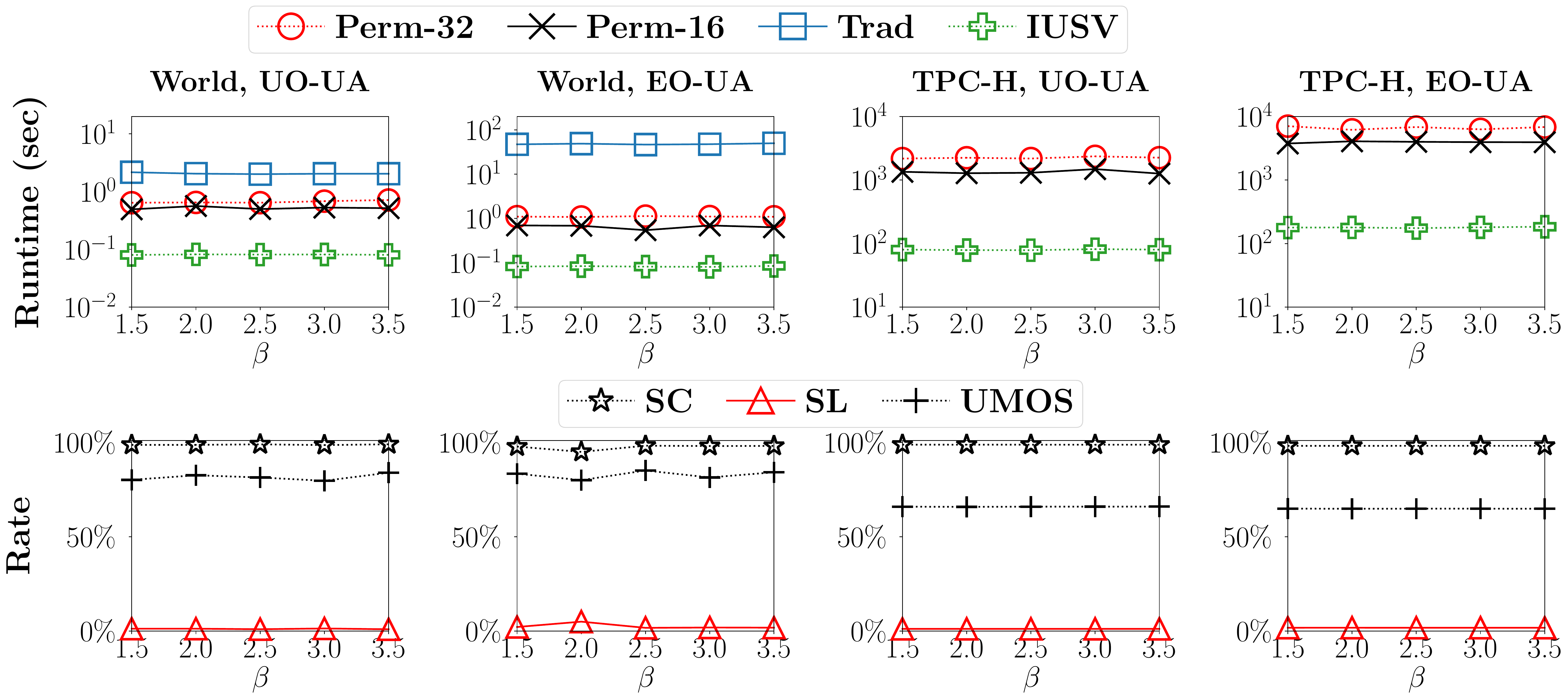}
    \caption{Effect of Zipfian Parameter $\beta$.}%
    \label{exp-fig:time_vs_beta}%
\end{figure}

In the unequal chance assignment of records to data owners (UA), the Zipfian distribution with parameter $\beta$ is used. \cref{exp-fig:time_vs_beta} shows the performance in the UO-UA and EO-UA settings when $\beta$ varies. The performance of all methods on both data sets is not sensitive to $\beta$. Each data owner can have up to one copy of a record. Once the number of copies for one record is determined (controlled by the Zipfian parameter $\alpha$ and constrained by the maximum copy of tuples $m$), the number of syntheses for a related tuple in the coalition set is largely determined, which is determinant to the performance of the proposed method. When the number of copies is low, due to the property of Zipfian distribution, there are only a small number of records with multiple copies. This leads to a small amount of tuples that yield a large number of syntheses. As a result, we observe that the SL algorithm invoking rate remains low. Whether a data owner holds most records of a table affects the individual data owners' Shapley values, but does not affect the computation cost much.

\begin{figure*}[t]%
    \centering
    \includegraphics[width=\linewidth]{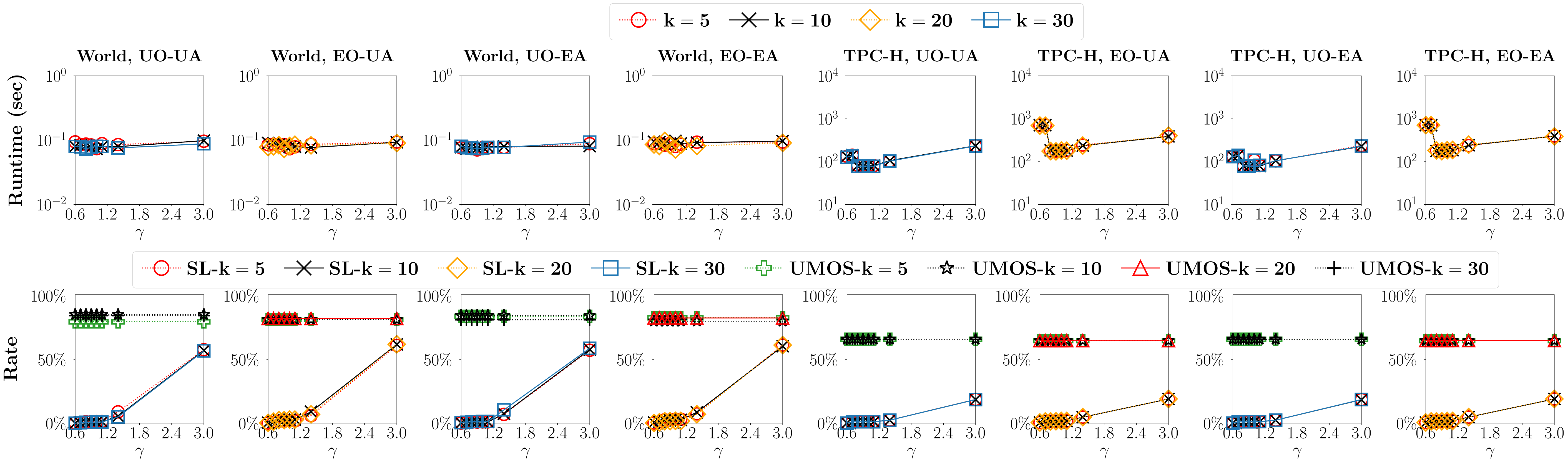}
    \caption{Effect of hyper-parameter $\gamma$ (Different Number of Data Owners $k$ Per Table)}%
    \label{exp-fig:time_vs_gamma_vs_owner}%
\end{figure*}

\begin{figure*}[t]%
    \centering
    \includegraphics[width=\linewidth]{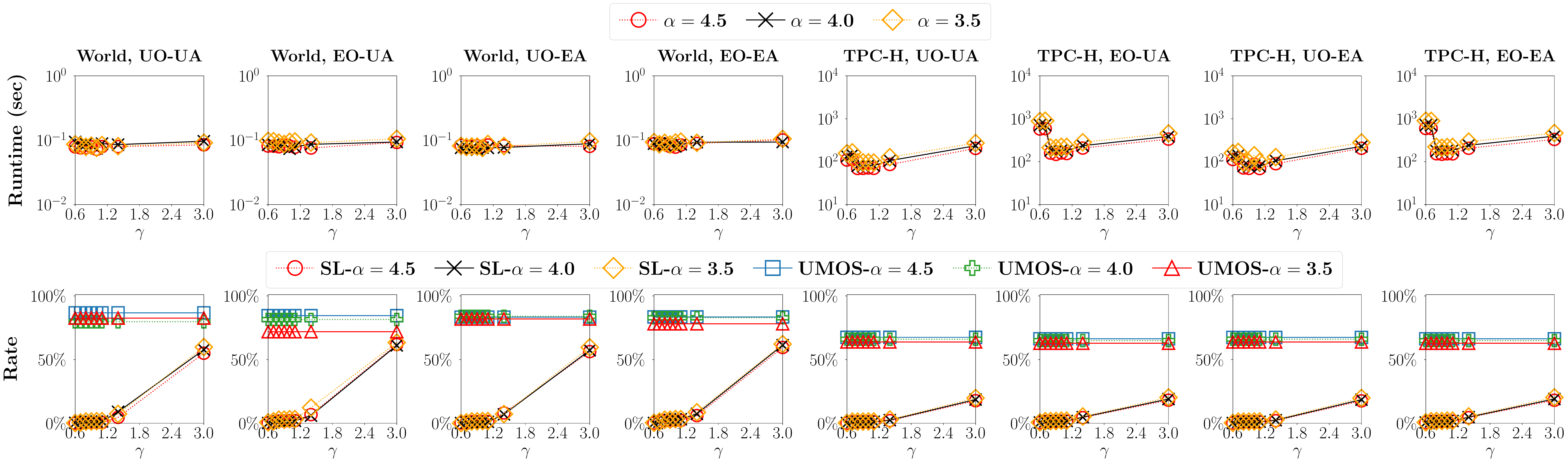}
    \caption{Effect of hyper-parameter $\gamma$ (Different Zipfian Parameter $\alpha$ values)}%
    \label{exp-fig:time_vs_gamma_vs_alpha}%
\end{figure*}

\begin{figure}[t]%
    \centering
    \includegraphics[width=1.05\linewidth]{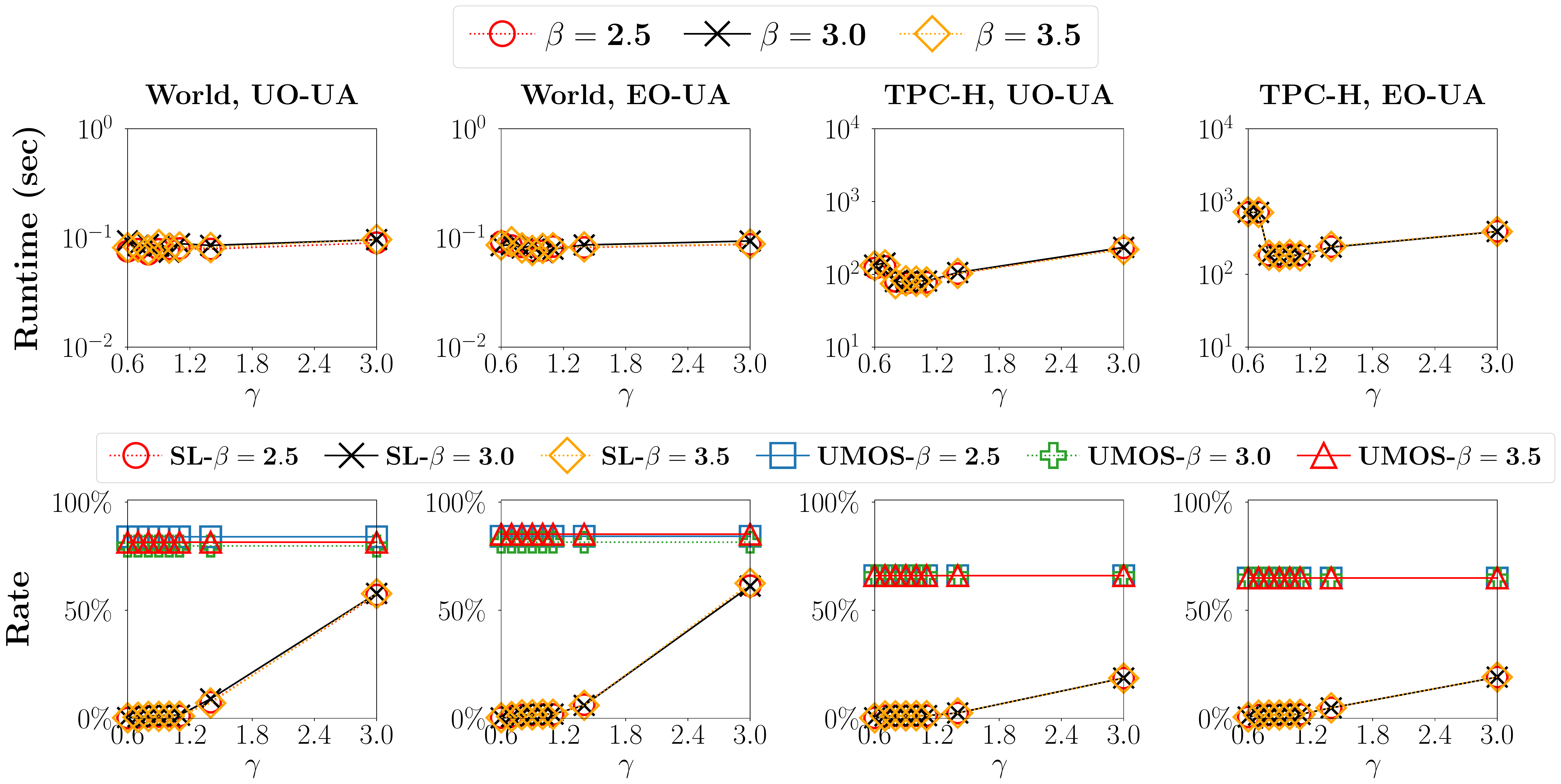}
    \caption{Effect of hyper-parameter $\gamma$ (Different Zipfian Parameter $\beta$ values)}%
    \label{exp-fig:time_vs_gamma_vs_beta}%
\end{figure}

\subsection{Effect of Hyper-parameter $\gamma$}

We investigate the performance of IUSV when the hyper-parameter $\gamma$ in \cref{algo:main} varies under the default data owner distribution. For the ease of reading, we only plot the SL rate, as the SC rate is 100\% minus the SL rate.  \Cref{exp-fig:time_vs_gamma_vs_owner,exp-fig:time_vs_gamma_vs_alpha,exp-fig:time_vs_gamma_vs_beta} show the results with respect to the number of data owners per table,  Zipfian parameter $\alpha$, and Zipfian parameter $\beta$, respectively.

When $\gamma$ increases, the runtime drops first when $\gamma$ is small, and then increases. This holds in all four settings. 
According to \cref{algo:main}, when $\gamma<1$, IUSV chooses the SC algorithm over the SL algorithm even when $\max(m_u, m_u m_{\bar{u}})$ is large, where in this case the SC algorithm is slow. This explains the initial drop.  When $\gamma$ keeps increasing, IUSV chooses the SL algorithm more and more, as shown by the SL rate. Since the SL algorithm may be more costly than the SC algorithm for tuples where the number of minimal synthesis owners is large, a good choice of $\gamma$ is close to 1. 

One important observation is that, when $\gamma$ is fixed, the performance of IUSV of various numbers of data owners and different values of $\alpha$ and $\beta$ is quite similar.  This observation suggests that setting $\gamma$ to a value around $1$ can achieve good performance for various settings.

\section{Conclusions}\label{sec:con}

As data sharing and integration becomes more and more important and popular, fair evaluation of data owners' contributions to a large data assemblage task remains challenging in computation.  While most of the existing work on Shapley value computation does not look into the specific characteristics in data assemblage, in this paper, we explore the decomposability of utility in data assemblage and formulate the independent utility assumption.  We discover that independent utility enjoys many applications and has a few interesting properties.  We develop fast and scalable algorithms for computing Shapley value under independent utility.  Our experimental results on real data sets show that our new approach is orders of magnitude faster than the conventional approaches.

Shapley value computation under independent utility opens a new direction for future work.  For example, it is interesting to explore whether the independent utility assumption can enable new opportunities for approximation of Shapley values
and incremental Shapley value computation on dynamic and streaming data.

\balance
\bibliographystyle{ACM-Reference-Format}
\bibliography{main/ref}

\end{document}